\documentclass[%
 reprint,
superscriptaddress,
 amsmath,amssymb,
 pre,
]{revtex4-2}

\usepackage{graphicx}% Include figure files
\usepackage{dcolumn}% Align table columns on decimal point
\usepackage{bm}% bold math
\DeclareMathOperator{\Tr}{Tr}
\usepackage{xcolor}

\newcommand{\modf}{\textcolor{blue}}

\begin{document}

%\preprint{APS/123-QED}

\title{Active nematic pumps}% Force line breaks with \\
\author{Ignasi Vélez-Cerón}
\altaffiliation{These authors contributed equally to this work}
\affiliation{Department of Materials Science and Physical Chemistry, Universitat de Barcelona, 08028 Barcelona, Spain}
\affiliation{Institute of Nanoscience and Nanotechnology, IN2UB, Universitat de Barcelona, 08028 Barcelona, Spain}
\author{Rodrigo C. V. Coelho}
\altaffiliation{These authors contributed equally to this work}
\affiliation{Centro Brasileiro de Pesquisas Físicas, Rua Xavier Sigaud 150, 22290-180 Rio de Janeiro, Brazil}
\affiliation{Centro de Física Teórica e Computacional, Faculdade de Ciências, Universidade de Lisboa, 1749-016 Lisboa, Portugal.}
\affiliation{Departamento de Física, Faculdade de Ciências, Universidade de Lisboa, P-1749-016 Lisboa, Portugal.}
\author{Pau Guillamat}
\affiliation{Institute for Bioengineering of Catalonia, The Barcelona Institute for Science and Technology, 08028 Barcelona, Spain}
\author{Marc Vergés-Vilarrubia}
\affiliation{Department of Materials Science and Physical Chemistry, Universitat de Barcelona, 08028 Barcelona, Spain}
\affiliation{Institute of Nanoscience and Nanotechnology, IN2UB, Universitat de Barcelona, 08028 Barcelona, Spain}
\author{Margarida Telo da Gama}
\affiliation{Centro de Física Teórica e Computacional, Faculdade de Ciências, Universidade de Lisboa, 1749-016 Lisboa, Portugal.}
\affiliation{Departamento de Física, Faculdade de Ciências, Universidade de Lisboa, P-1749-016 Lisboa, Portugal.}
\author{Francesc Sagués}
\author{Jordi Ignés-Mullol}
\affiliation{Department of Materials Science and Physical Chemistry, Universitat de Barcelona, 08028 Barcelona, Spain}
\affiliation{Institute of Nanoscience and Nanotechnology, IN2UB, Universitat de Barcelona, 08028 Barcelona, Spain}

%\noaffiliation%\collaboration{MUSO Collaboration}

\date{\today}% It is always \today, today,

\begin{abstract} %Max 200 words
\textbf{Abstract}. Microfluidics involves the manipulation of flows at the microscale, typically requiring external power sources to generate pressure gradients. Alternatively, harnessing flows from active fluids, which are usually chaotic, has been proposed as a paradigm for the development of micro-machines. Here, by combining experimental realizations and simulations, we demonstrate that the addition of triangular-shaped obstacles into an active nematic gel can locally break the fore-aft symmetry of active turbulence and stabilize flow fields with self-pumping capabilities. The proposed strategy has enabled us to generate wall-free and self-powered microfluidic systems capable of both cargo transport and mixing along with the downstream flow. We analyze the performance of these active pumps, both isolated and within cooperative ensembles in terms of their output velocity and hydrostatic pressure buildup. Finally, we demonstrate strategies to incorporate them into specifically designed microfluidic platforms to advantageously tailor the geometry of active flows. Our results reveal new possibilities for leveraging the self-organized mechanodynamics of active fluids.\\

\textbf{Significance statement}. Lab-on-a-chip devices allow multi-sample chemical analysis and complex in-situ microfabrication in an integrated platform where fluids are propelled and controlled at submillimeter scales. While these systems are typically hampered by the need to rely on external power sources, we demonstrate here an alternative strategy using active fluids, composed of self-propelled units that normally develop disordered flows. We work with an active nematic fluid where we embed lattices of triangular hydrogel micro-columns, which transform the turbulence into controllable flow patterns, effectively harnessing the local activity and leading to the emergence of active pumps that can be arbitrarily combined to internally adjust the pressure or flow rate in an integrated microfluidic device.

\end{abstract}

\maketitle

%\tableofcontents

%Supplementary figueres and movies

\makeatletter
\newcommand{\manuallabel}[2]{\def\@currentlabel{#2}\label{#1}}
\makeatother

%\manuallabel{SFig:director}{S1}
\manuallabel{SFig:profiles}{S1}
\manuallabel{SFig:elements}{S2}
\manuallabel{SFig:small}{S3}
\manuallabel{SFig:random}{S4}
\manuallabel{SFig:Vij}{S5}
\manuallabel{SFig:jumps}{S6}
\manuallabel{SFig:mixing}{S7}
\manuallabel{SFig:peclet}{S8}
\manuallabel{SFig:injection_sim}{s9}
\manuallabel{SFig:channel}{S10}
\manuallabel{SFig:circles}{S11}
\manuallabel{SFig:circulation_sim}{S12}

\manuallabel{SMov:experiments}{S1}
\manuallabel{SMov:simulations}{S2}
\manuallabel{SMov:tracers_exp}{S3}
\manuallabel{SMov:tracers_sim}{S4}
\manuallabel{SMov:mixing}{S5}
\manuallabel{SMov:circuit}{S6}
\manuallabel{SMov:tracers_circuit_sim}{S7}
\manuallabel{SMov:inversion_exp}{S8}
\manuallabel{SMov:inversion_sim}{S9}

\section{Introduction}

Microfluidic systems have become fundamental tools in modern science and technology, enabling precise manipulation of fluids at the microscale for applications ranging from chemical synthesis and microfabrication to biomedical devices \cite{Battat2022}. While these systems have been extremely useful in a wide range of scientific and industrial fields, both their geometrical constraints and the fact that they depend on external pressure sources impose limitations on systems design, downscaling, and autonomy.

A transformative approach to microfluidics could be explored by transitioning from traditional to active microfluidic systems, leveraging the flows generated by self-powered and self-organized biological or bioinspired active fluids \cite{dileonardo2010,Thampi2016,ray2023}, as well as their intrinsic control mechanisms, refined during millions of years of natural evolution. Constituted by both synthetic particles \cite{Bricard2013} and self-propelled entities of biological origin, such as bacteria \cite{Dombrowski2004,Wensink2012,Zhou2014}, mamallian cells \cite{Lu2023,Fletcher2021}, or cytoskeletal proteins  \cite{Sanchez2012,Murrell2021}, active fluids efficiently convert chemical energy into sustained motion, typically in the form of seemingly disordered flows \cite{ramaswamy10,marchetti2013,needleman17,Sagues23}. 
%The intrinsic length and time scales emerging in active flows are crucial organizers of biological systems across scales, ranging from animal groups, to bacterial colonies or cellular tissues \cite{needleman17}. In cells, flows by the cytoskeletal networks ensure proper nutrient mixing and transport \cite{Lu2023}, and provide with the forces for cells' deformation and locomotion \cite{Fletcher2021}. In tissues, collective active flows govern the mechanodynamics of crucial processes for life, such as wound healing~\cite{Tetley2019} or morphogenetic events during development~\cite{MaroudasSacks2020}. 
Understanding how to regulate the flows of these active fluids \cite{wu2017,shankar24} can thus be a handle to  harnessing their mechanodynamics into well-organized flow patterns for the development of new microfluidic technologies.

Within the state-of-the-art active systems, synthetic active gels stand out as exceptional materials \cite{Prost15}. They combine the best attributes of biology and colloidal sciences, offering not only efficient self-organized behaviors and adaptability but also robustness and ease-of-manipulate. In particular, active nematic (AN) gels, endowed with orientational order, provide additional unique properties that enhance their functionality and controlability \cite{doostmohammadi2018active}.  In this regard, some promising steps toward controlling ANs have been realized in the past few years. For instance, it has been experimentally demonstrated that spatial confinement within channels or rings can lead to an effective rectification of turbulent flows in active nematic preparations, both in 3d  and in 2d, resulting in regimes with effective transport \cite{wu2017,opathalage19, Hardouin2019,hardouin2020}. Flow control of active fluids can also be exerted using interfacial conditioning \cite{guillamat2016,sunyer2016}, or light patterning \cite{zhang2021}. The symmetry breaking that leads from isotropic active turbulence to net active transport needs a mechanism to locally set both the orientation and the handedness of the flows. This has been typically addressed by using ratchet-like boundaries \cite{caballero2015,wu2017,hardouin2020}, which have enabled to rectify the otherwise random rotation of inclusions embedded in an active fluid \cite{dileonardo2010,sokolov2010,ray2023}. In spite of these developments, we still lack a versatile strategy to unleash the full microfluidic potential of active nematic flows.

\begin{figure*}[htb]
\includegraphics[width=\textwidth]{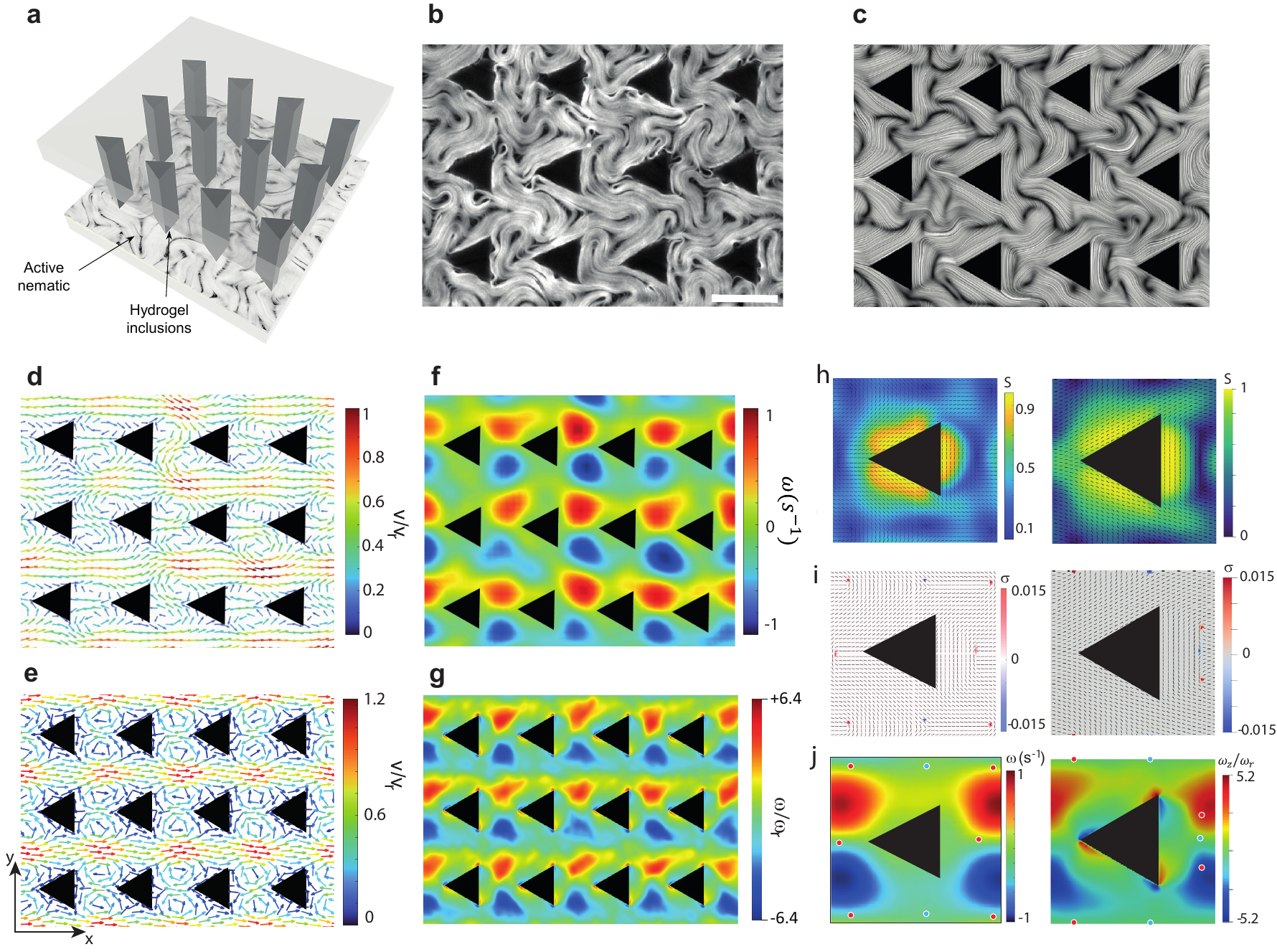}
\caption{\textbf{Active pumping with a square array of triangular obstacles.} \textbf{a} An array of hydrogel columns with triangular cross section is  photopolymerized in-situ in the aqueous phase of an AN flow cell. Scale bar 100 $\mu$m. \textbf{b} Fluorescence image showing the adaptation of the AN to the presence of the obstacle array (see Movie \ref{SMov:experiments}). \textbf{c} Simulated system (see Movie \ref{SMov:simulations}). The instantaneous director field is rendered using the Line Integral Convolution filter. Diagrams \textbf{d} and \textbf{f} correspond to the analysis of the experimental flow field, while \textbf{e} and \textbf{g} correspond to the analysis of the simulated flow field. Experimental images correspond to the central part of a lattice containing 6 rows and 9 columns in total, while the simulated system, also only partially shown, contains 6 rows and 30 columns with periodic boundary conditions in both directions. Time-averaged velocity (\textbf{d}, \textbf{e}) and vorticity (\textbf{f}, \textbf{g}) fields. $v_r$ is the average speed and, in the simulations, $\omega_r = v_r/D_{def}$, where $D_{def}$ is the average distance between defects, all measured in the absence of obstacles. 
%\textbf{h}, \textbf{i} X-component of the velocity fields along the horizontal center line between rows of obstacles (averaged in time and for different \attn{rows?}), normalized with $v_r$. Panel \textbf{i} corresponds to a simulated system with 30 triangles along the horizontal direction, inside a closed box with no-slip boundary conditions. 
%\textbf{h} Simulated average horizontal flow velocity relative to $v_r$ as a function of the size of the equilateral triangle, $H$, for three different normalized activity coefficients, $\zeta D_{def}/(v_r \eta)$, \modf{and for $H/L = 0.55$}. Here, $\eta$ is the two-dimensional active fluid viscosity. Simulations are performed in one lattice cell, using periodic boundary conditions. Data are collapsed by rescaling $H$ with $D_{def}$.  \textbf{i} Diagram of the normalized average horizontal velocity, $v_x/v_r$, \modf{constructed from simulations,} as a function of the normalized triangle size, $H/D_{def}$, and triangle size relative to cell size, $H/L$. Simulations are performed with $\zeta D_{def}/(v_r \eta) = 7.84$.}
\textbf{h}-\textbf{j} Average active nematic fields around triangular obstacles. Left column: experiments; right column: simulations. Conditions are the same as in the other panels in this figure. The field of view in simulations is exactly one cell in the periodic lattice, while in experiments the field of view is slightly larger than the unit cell. \textbf{h} Scalar nematic order parameter, $S$, and director field. \textbf{i} Topological charge density, $\sigma$, and director field. \textbf{j} Vorticity field with hotspots from the topological charge density map highlighted.
}
\label{Fig:setup_results}
\end{figure*}

\begin{figure}[t]
\includegraphics[width=0.4\textwidth]{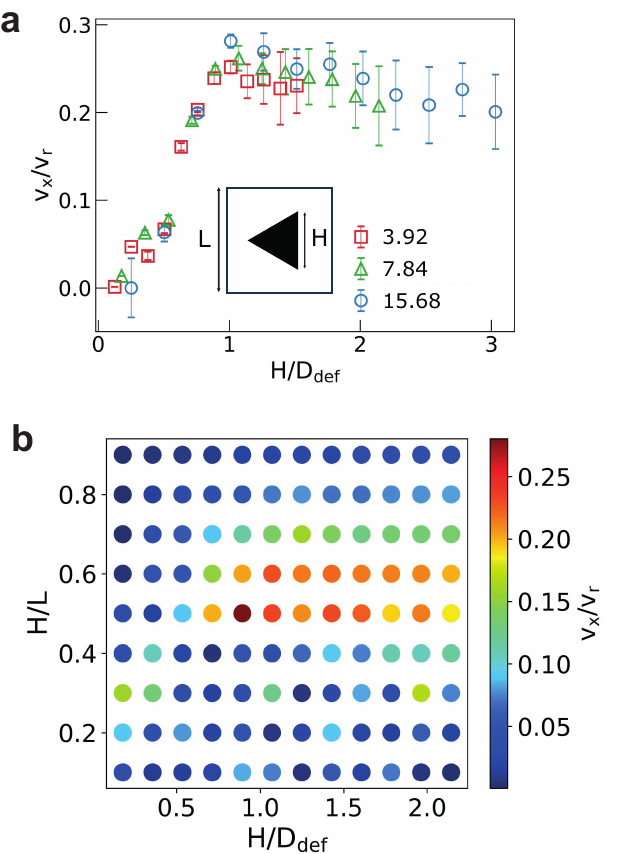}
\caption{\textbf{Optimal active pump geometry.} \textbf{a} Simulated average horizontal flow velocity relative to $v_r$ as a function of the size of the equilateral triangle, $H$, for three different normalized activity coefficients, $\zeta D_{def}/(v_r \eta)$, and for $H/L = 0.55$. Here, $\eta$ is the two-dimensional active fluid viscosity. Simulations are performed in one lattice cell, using periodic boundary conditions. Data are collapsed by rescaling $H$ with $D_{def}$.  \textbf{b} Diagram of the normalized average horizontal velocity, $v_x/v_r$, constructed from simulations, as a function of the normalized triangle size, $H/D_{def}$, and triangle size relative to cell size, $H/L$. Simulations are performed with $\zeta D_{def}/(v_r \eta) = 7.84$.}
\label{Fig:phases}
\end{figure}

Here, we address this gap by demonstrating that obstacles with well-defined geometrical asymmetry can lead to the phenomenon of active pumping. In particular, we take advantage of a recent protocol to polymerize biocompatible hydrogel structures directly into a fully-developed active nematic gel \cite{velez2024} to imprint lattices of triangular obstacles that, for sizes comparable to the intrinsic active length scale, lead to hydrostatic pressure buildup and enhanced unidirectional flows. Combining experiments and numerical simulations we show that the observed phenomenon is a cooperative effect emerging from a lattice of counter-rotating vortices that result from the presence of obstacles with broken fore-aft symmetry within the active nematic fluid. By characterizing the interplay between flow patterns and pressure gradients generated within the active nematic fluid, we demonstrate that arrays of triangular obstacles can lead to efficient active nematic pumping through the creation of virtual flow channels. The absence of physical confinement walls makes this system an ideal candidate for large-scale active microfluidic transport and mixing of colloidal cargo, or to locally adjust the geometry of the active fluid~\cite{D2SM00988A}. \modf{Confinement-free effects also distinguish our scenario from that of the pumping of a threshold-less active nematic suggested by Green et al. \cite{Green2017}. In this case, flows are triggered to balance the force density created by geometrical distortions. }

\begin{figure*}[htb]
\includegraphics[width=\textwidth]{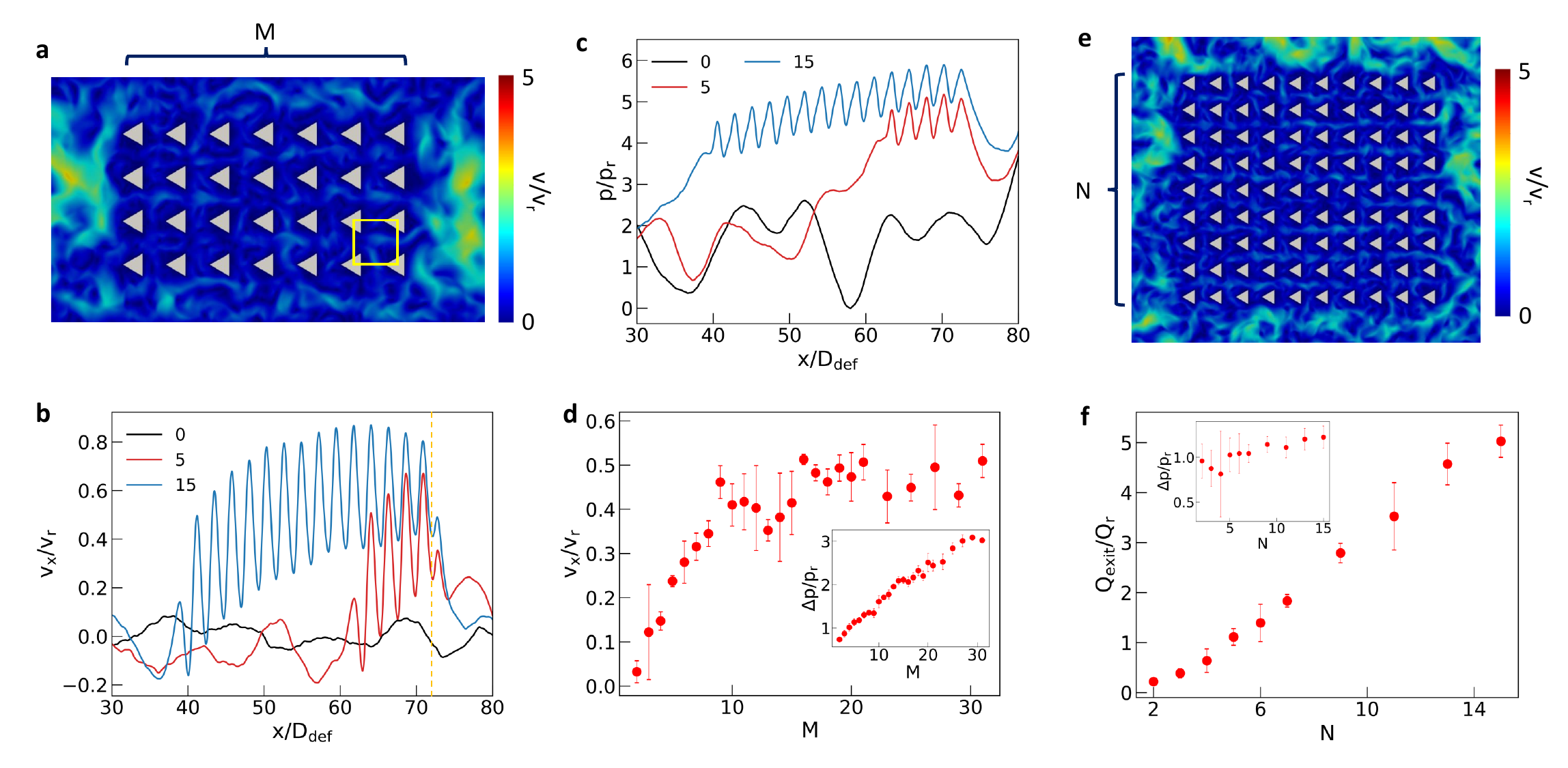}
\caption{\textbf{\modf{Active pumps systems}}.
\textbf{a} \modf{Simulated study of the performance of an array of pumps as we change the number of columns, $M$. The number of rows is kept constant at 4 (3 flow channels). The yellow square represents a single pump in the region between four triangles.} The size of the simulated domain is $ 83.2\, D_{def} \times 13.7\, D_{def}$. The simulation conditions are $H=1.25 D_{def}$, $L=2.3 D_{def}$, $\zeta D_{def}/(v_r \eta) = 7.84$ and closed box with no-slip boundary conditions, are the same in all cases.  \textbf{b} Velocity field along the central lines between rows of triangles (averaged for the three rows and with time) \modf{for different number of columns, including $M=0$} (closed box without triangles) for comparison. The orange dashed vertical line indicates the exit of the pump region. \textbf{c} For the same simulations as in (\textbf{b}), we monitor the pressure field \modf{for three different values of $M$}. Here we plot the normalized pressure profile, averaged for the three rows and with time, and normalized with $p_r=\rho_0 v_r^2$, where $\rho_0$ is the average density. \textbf{d} Average normalized horizontal velocity at the output of the \modf{pump array for different number of columns}. The inset indicates the corresponding pressure difference across the pump system. The error bars represent the standard deviation for different random initial conditions.  \textbf{e} \modf{Array of pumps where we keep the number of columns fixed at 10 and we vary the number of rows, $N$.} The size of the simulated domain is
$\rm 37.7\, D_{def} \times 43.4\, D_{def}$. The size of the unit cell and the simulation conditions are the same as above. \textbf{f} Average flow rate at the output for different \modf{number of rows}, relative to the reference flow rate $Q_r=v_r D_{def}$. The inset depicts the pressure difference across the pump system.
}
\label{Fig_pumps}
\end{figure*}

\section{Results}

\subsection{Creation of active nematic pumps}

We photopolymerize square lattices of hydrogel columns with triangular cross section inside an AN flow cell \cite{velez2024}. The columns are attached to the hydrophilic glass plate and extend across the AN towards the thin oil layer on the complementary hydrophobic plate (see Fig. \ref{Fig:setup_results}a and Methods).  As a result, a lattice of rigid triangular obstacles is suddenly embedded within an AN layer in the active turbulent regime. Such intervention prompts the adaptation of its orientational and velocity fields to the new in-plane boundary conditions (Fig. \ref{Fig:setup_results}b). Upon polymerization, filaments in irradiated regions are frozen in non-specified orientations. This seems to have no impact either on the surrounding orientational or velocity fields of the embedded AN, which organize similarly around all obstacles.

Both experiments and continuum hydrodynamic simulations (Fig. \ref{Fig:setup_results}c; see Methods) show that the active flow patterns organize around the array of obstacles leading to the formation of virtual flow channels. The velocity field is oriented, on average, along the space between rows of obstacles, antiparallel to the orientation of the triangles in the array (left-to-right in the images), as shown in Fig. \ref{Fig:setup_results}d and Fig. \ref{Fig:setup_results}e. This is different from active nematic flows confined in channels, which can exhibit net transport of the normally turbulent-like flows, but without a predefined flow polarity \cite{wu2017,Hardouin2019}.  
%In the diagrams, we have represented the velocity field normalized with the average speed in the absence of obstacles, $v_r$. 
In particular, we observe that active flows organize, on both sides of the frontal tip of each triangular obstacle, as a pair of counter rotating vortices, which pump the AN layer. This is clearly highlighted in Fig. \ref{Fig:setup_results}f and g, where we represent the average vorticity field of the active flows. 

Active nematic flows are know to be organized by self-propelling +1/2 defects, which form through spontaneous unbinding of opposite $\pm1/2$ pairs \cite{doostmohammadi2018active}, and are also injected into the system by the active boundary layers that form along the obstacle walls \cite{Hardouin2022}. We have measured the time-averaged orientational field around triangular obstacles, both from experiments and numerical simulations (Fig. \ref{Fig:setup_results}h), and we have computed the corresponding topological charge density field, which is correlated with the local presence of defects (Fig. \ref{Fig:setup_results}i). We observe that the order parameter is maximal on the obstacle walls, where the nematic director adopts, on average, planar alignment, and a pattern of aligned $\pm1/2$ defects emerges along the virtual flow channels, with +1/2 defects pushing the flow and $-1/2$ defects being advected by it (Fig. \ref{Fig:setup_results}j).

Our observations are consistent with earlier simulations that reported a steady vortex lattice forming in AN confined in an array of astroid-shaped obstacles \cite{schimming2024}. In our case, the flows emerging from counter-rotating vortex pairs leads to net transport in a direction that is predefined by the orientation of the obstacles in the array. Inside the virtual channels, flow speed is non monotonous, being higher between vortex pairs. Overall, this leads to a unidirectional pumping mechanism. (Fig. \ref{SFig:profiles}). A similar flow rectification effect has been recently proposed using numerical simulations with half-astroid-shaped obstacles \cite{schimming2024b}, consistent with the expected ratchet effect in the presence of asymmetric obstacles. 

The reported pumping mechanism depends on the number, size, shape, and spatial distribution of the triangles in the array.  A stable vortex lattice emerges only if at least a 2 by 2 square lattice of obstacles is present (Fig. \ref{SFig:elements}). We consider this arrangement to be the minimal system to achieve active pumping. In Fig. \ref{Fig:setup_results}, both in experiments and in numerical simulations, the size and spacing of the triangles are chosen to be equal to the average spacing between defects in the absence of obstacles (see early stages of Movie S1), $D_{def}$, which we have determined to be of the order of 70 $\mu$m in experiments and 56 lattice units in simulations.

Using numerical simulations, we next study the effect of these two parameters. In Fig. \ref{Fig:phases}a, we show the average normalized horizontal velocity as a function of the normalized size of the triangles, $H/D_{def}$, for a fixed spacing $L=2.3 D_{def}$. Note that, in the absence of AN pumping effects, the flows are turbulent-like and therefore, the net velocity is zero in any direction. We observe a quick increase of the pumping efficiency with obstacle size, until $H \simeq D_{def}$, slowly decreasing for larger obstacles. We have performed simulations for different activity coefficients (related to the concentration of adenosine triphosphate in experiments), and find that the results remain invariant as the velocity and spatial dimensions are rescaled with the intrinsic velocity and active length scale in the absence of obstacles at each activity \cite{giomi2015,doostmohammadi2018active}.

We next use simulations to explore, for a constant activity, the optimal geometry of the square lattice of triangles, in terms of their size and spacing. The resulting phase diagram, shown in Fig. \ref{Fig:phases}b, indicates that the optimal arrangement corresponds to $H \simeq D_{def}$ and $L\simeq 2D_{def}$, which we have used in the experiments and simulations in Fig. \ref{Fig:setup_results}, and will be used henceforth by default. Our experimental technique allows to fabricate more compact active pump arrangements, as shown in Fig. \ref{SFig:small}, thus demonstrating the capabilities for miniaturization of these devices.

The pumping effect is also severely affected by the lattice regularity, as analyzed with simulations in Fig. \ref{SFig:random} for lattices of triangles with different degrees of disorder with respect to the perfect square lattice arrangement. We observe that a 
random lateral displacement of the triangles' center  25\% with respect the square lattice nodes halves the efficiency of the AN pumping effects, which are completely lost when lateral displacement is higher than 50 \%.

\subsection{Active pump systems}
After determining the optimal geometric features of basic AN pumps, we consider the design of more intricate microfluidic systems that include multiple AN pump units (Fig. \ref{SFig:elements}). \modf{In particular, we have studied the performance of square lattices of triangles for different number of columns and rows (Fig. \ref{Fig_pumps}). }

%(Fig. \ref{Fig_pumps}) for different number of columns (active pumps in series, Fig. \ref{Fig_pumps}a-d) and rows (active pumps in parallel, Fig. \ref{Fig_pumps}e,f). 

\modf{By keeping the number of rows fixed, we can study the effect of increasing the number of columns (Fig. \ref{Fig_pumps}a-d), which, in the language of hydraulics, is the  equivalent of concatenating AN pumping units in series. We observe} that the average flow speed increases with the number of columns (Fig. \ref{Fig_pumps}b), until reaching a maximum value at about 10 pumps in series (Fig. \ref{Fig_pumps}d). We also consider the concurrent profiles of the hydrostatic pressure within the active fluid along the flow channels, and we observe that it increases steadily with the number of pumping elements in series, even when the average speed has reached its saturation value (Fig. \ref{Fig_pumps}c and inset in Fig. \ref{Fig_pumps}d). In our active pump system, pressure gradients are positively correlated with the flow velocity. This is contrary to flow through a microfluidic channel or through a porous medium, where velocity is negatively correlated with the driving pressure gradients, even in the case of pressurized active fluids \cite{keogh2024}. In classical hydraulics, a system of ideal pumps in series yields a flow velocity that does not change with the number of pumps and a pressure that builds up linearly with the number of elements~\cite{karassik2008pump}. Interestingly, our active pumps behave as ideal hydraulic pumps when the number of elements in series is larger than 10, as observed in Fig. \ref{Fig_pumps}d. The impossibility of a continuously increasing speed in a \modf{sequence} of active nematic pumps can be justified by the fact that energy injection is finite. Indeed, as speed increases, viscous dissipation will eventually match energy injection and the input and output speed of subsequent pumping elements will remain the same.

\begin{figure}[htb]
\includegraphics[width=0.45\textwidth]{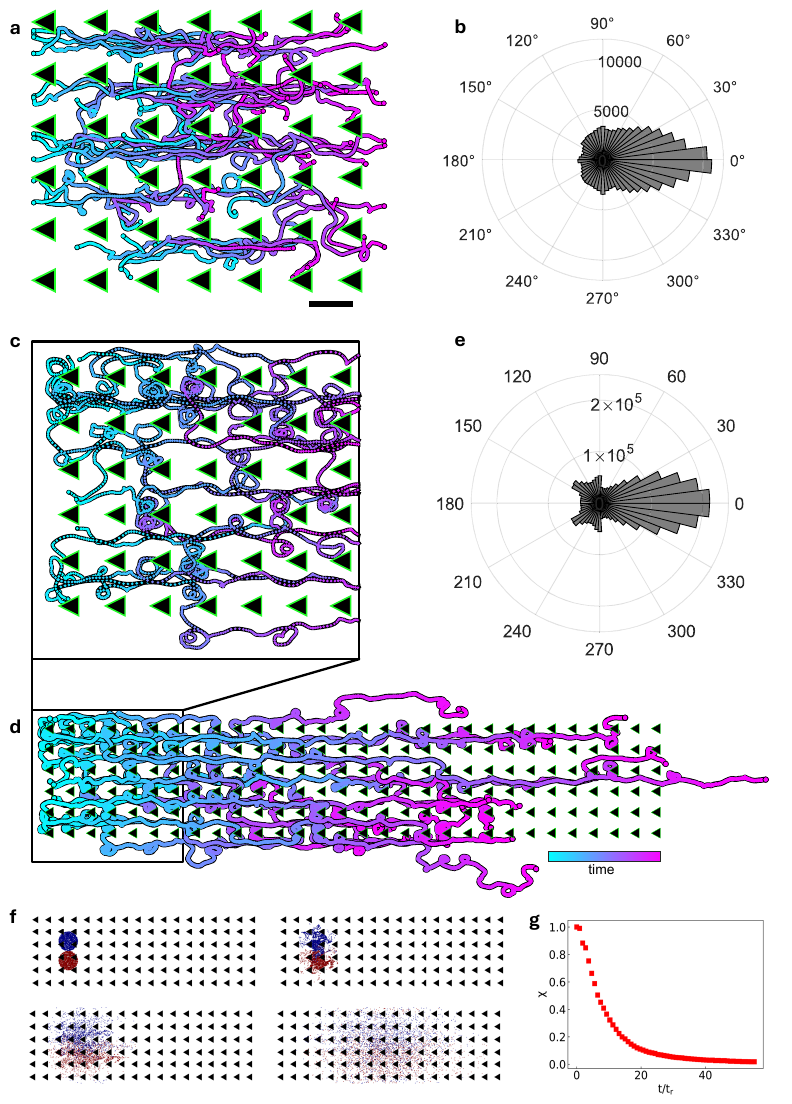}
\caption{\textbf{Transport and mixing with active pumps}. \textbf{a} Experimental trajectories of colloidal particle tracers being advected by active flows within a lattice of triangules (see Movie \ref{SMov:tracers_exp}). Scale bar 100 $\mu$m. \textbf{b} Distribution of tracer velocity orientations for the same experiment. Data accumulated with 1579 trajectories. \textbf{c,d} Simulated trajectories of tracers being advected by active flows (see Movie \ref{SMov:tracers_sim} and the corresponding distribution of velocity orientations (\textbf{e}). Simulation conditions are $H=1.25 D_{def}$, $L=2.3 D_{def}$, $\zeta D_{def}/(v_r \eta) = 7.84$. and periodic boundary conditions in both directions. \textbf{f} Simulations with the mixing of two species being advected by active flows (see Movie \ref{SMov:mixing}). The reference time is $t_r=D_{def}/v_r$. The four figures correspond to the following times respectively: 0, 6.2$t_r$, 18.8$t_r$, 54.6$t_r$.
\textbf{g} Temporal evolution of the mixing parameter, $\chi$, defined in the main text.}
\label{Fig:mixing}
\end{figure}

\begin{figure*}[htb]
\includegraphics[width=\textwidth]{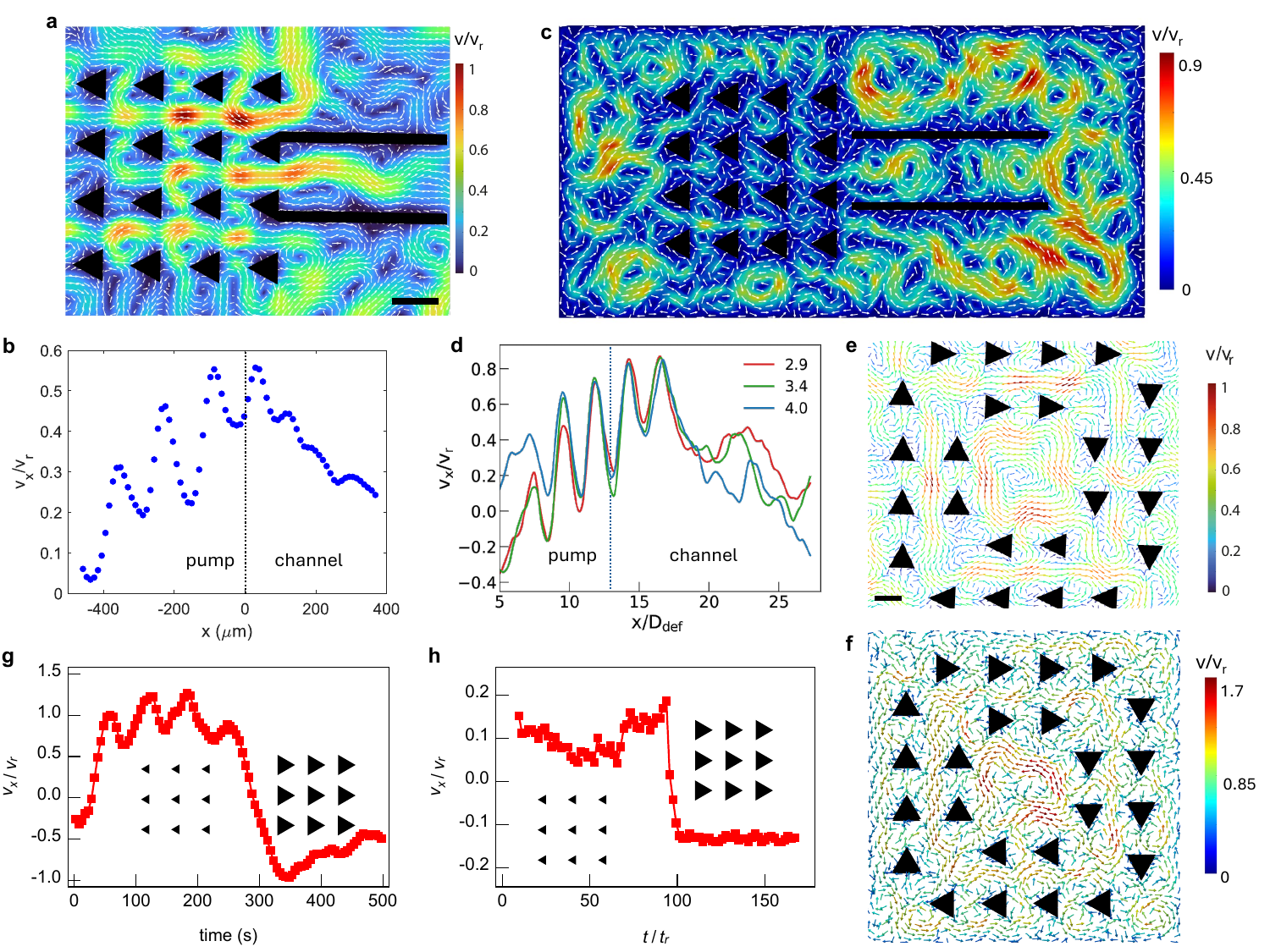}
\caption{\textbf{Microfluidic applications of active nematic pumping}. \textbf{a} Experiments where the output of an active pump system is injected into a channel with smooth walls of width $L_c \simeq 3 D_{def}$, with $D_{def}\simeq 70\mu$m. Triangular obstacles and channel boundaries are overlaid on the map of the velocity field. Scale bar 100 $\mu$m.  \textbf{b} Simulation of an active pump system injecting into a channel of width $L_c=2.3 D_{def}$ and activity $\zeta D_{def}/(v_r \eta) = 7.84$. We perform simulations in a closed box of dimensions %$24 \times 6$ cells (or 
$\rm 54.9\, D_{def} \times 13.7 \, D_{def}$, with no-slip boundary conditions on all walls. 
\textbf{c} Horizontal velocity (averaged over the central $50\,\mu$m along the channel) relative to $v_r$ for the experiments in \textbf{a}.
\textbf{d} Velocity field along the central part of the channel, averaged in time and for the different initial conditions in simulations, for three different normalized channel widths $L_c/D_{def}$. \textbf{e} Experimental time-averaged velocity field in the presence of a closed circuit of elementary active pumps. Scale bar $100 \mu$m (see also Movie \ref{SMov:circuit}). %
\textbf{f} Simulations of larger scale flows guided by a lattice of triangles. Counterclockwise flow circulation is guided by the active pump system. %The white and the grey trajectories illustrate the path followed by two tracers and the white rectangle indicates the path used to calculate the circulation. 
The system, with dimensions %$16 \times 10$ cells (or% 
$\rm 36.6\, D_{def} \times 18.3\,D_{def}$ (larger than the region shown in the panel), is in a closed box with solid walls at the borders. See also movie \ref{SMov:tracers_circuit_sim}.  \modf{\textbf{g} and \textbf{h} are an example of pump reversal. Average relative horizontal velocity profile inside the flow channels corresponding to experiments (\textbf{g}) and simulations (\textbf{h}) for an array of  triangular obstacles where new, larger obstacles facing in the opposite direction are imprinted on top of the original ones. In experiments, the square array has 70 $\mu$m spacing, triangles have an initial side length 27 $\mu$m and final side length 50 $\mu$m. Obstacles are created at t = 20 s and modified at t = 267 s. In simulations, array spacing is $2.3 D_{def}$, triangles have an initial side length $0.8 D_{def}$ and final side length $1.6 D_{def}$. Simulations are performed on a 5 x 5 array with periodic boundary conditions.} 
}
\label{Fig:channel}
\end{figure*}

\modf{Next, we analyze the performance of a lattice of active pumps when we increase the number of rows, keeping the number of columns fixed at 10, i.e., 9 pumps in series (Fig. \ref{Fig_pumps}e,f).  In the language of hydraulics, this corresponds to changing the number of pumps in parallel, $N$. The output velocity, which does not change with $N$, is the one expected with 9 pumps in series.  We will thus characterize each pump system with its average flow rate, which we find to be proportional to $N$, just as it would be for a system of ideal hydraulic pumps in parallel (Fig. \ref{Fig_pumps}f). We have also monitored the average pressure drop across the system, which we find to be constant, independent of $N$. These observations indicate that neighboring pumping elements along the same column operate independently, as a system of ideal pumps, despite the fact that the separation between adjacent flow channels is permeable.} This favors not only net transport but also matter exchange, as we show in the next section. 

\subsection{Micro-fluidic transport and mixing by active nematic pumps}
Next we demonstrate how to harness active flows to achieve effective microfluidic transport and mixing. This is implemented by locally imprinting our self-pumping system that, being realized along virtual channels, readily enables exchange of matter across their discontinuous walls. Both experiments (Fig. \ref{Fig:mixing}a,b) and simulations (Fig. \ref{Fig:mixing}c,d) reveal that passive colloidal particles are advected downstream by active flows, despite following erratic trajectories. Indeed, particles can be trapped in a vortex or they can switch to neighboring channels in the parallel pump system, but with a net average displacement clearly oriented downstream (Fig. \ref{Fig:mixing}b,d). A detailed statistical study of particle velocities (Fig. \ref{SFig:Vij}) show that they reach a steady average value along the channels with the orientation set by the triangular obstacles. Conversely, the transversal ($V_y$) component has a vanishing average value, consistent with no net transversal flow, but the average value of $|V_y|$ is finite, indicating an active exchange of particles across the virtual channel walls. 

On the other hand, channel permeability represents an enormous advantage with respect to classical microfluidic devices, and allowed us to achieve 
enhanced downstream mixing. We tested this concept by simulating an active pump system where adjacent circular ensembles of two distinctive colloidal species (red and blue in Fig. \ref{Fig:mixing}f) are advected by active flows. We observe that, while the center of mass of the ensemble moves downstream, mixing of the transported particles takes place, both in the longitudinal direction within channels and between neighboring channels. In fact, most particles move to a different channel during the process, as demonstrated in Fig. \ref{SFig:jumps}, where we show that the probability of a particle remaining in its original channel drops exponentialy with time. The effectiveness of mixing is exemplified in Fig. \ref{Fig:mixing}g with the temporal evolution of a mixing parameter, $\chi = \langle (N_A-N_B)^2 \rangle /\text{max} \langle (N_A-N_B)^2 \rangle$, where $N_A$ and $N_B$ are the number of particles of the two species at each position (within a square of side half the cell size). Perfect mixing corresponds to $\chi = 0$. We observe a drop from $\chi = 1$ (unmixed) to $\chi \simeq 0.1$ (near perfect mixing) after the ensemble has crossed $\sim 11$ pumping elements. Although  mixing is more efficient in fully turbulent active flows performing free of lateral constraints, our pumping system combines mixing with downstream transport, something that cannot be achieved through a lattice of isotropic obstacles or inside a channel with smooth walls (Fig. \ref{SFig:mixing}). %The proposed strategy harness the self-sustained flows generated by active nematics into active fluid pumping, possessing necessary qualities for autonomous transport and mixing in future bioenabled microfluidic devices. 
To put the performance of our active pumps in the context of microfluidic mixing \cite{Nguyen2005}, we have computed a Péclet number, $Pe= v L /D$, which compares particle transport by advection and by diffusion \cite{Squires2005}. To this end, we have extracted an effective diffusion coefficient along the transverse direction and used, as advection velocity, the average speed along the virtual channels (Fig. \ref{SFig:peclet}). In experiments, we have $\langle|v_x|\rangle \simeq 1\,\mu$m/s, $L \simeq 100\,\mu$m, and $D_{eff} = 15.2\,\mu m^2/s$, which yields $Pe\simeq 6.6$. In simulations, we have $\langle|v_x|\rangle \simeq 0.0042\,l.u.$, $L \simeq 128\,l.u.$, and $D_{eff} = 0.077$ l.u., which gives $Pe\simeq 7.0$, consistent with experiments. $Pe$ values close to unity, such as the ones obtained here, indicate that diffusion and advection times are comparable for the transport of dispersed particles. Note that $D_{eff}$ is several orders of magnitude larger than $D$, the thermal diffusion coefficient, as the bulk viscosity of the AN layer is $\eta\simeq 1$ Pa·s \cite{velez2024}. The observed diffusion enhancement is an example of active Taylor dispersion \cite{Squires2005}, a phenomenon amply studied in passive liquids, but that has been addressed only recently for active fluids \cite{Peng2020, alvim2024solutedispersionpreturbulentconfined}.

\subsection{Integrating active pumps into microfluidic systems}
The versatility of our active nematic pumping system enables its incorporation anywhere and at anytime within an active fluid layer. As an example, in Fig. \ref{Fig:channel}a we demonstrate the efficacy of our control strategy by directing the flow from the outlet of an AN-pumped virtual channel into a channel confined by physical walls. It is known that flow inside a channel narrower than the active length scale can lead to structured flow patterns and even to net transport, although the resulting flow orientation is randomly selected upon symmetry breaking \cite{hardouin2020}. In the present case, the orientation of the flow is determined by the active pumping system. Here, both triangular obstacles for pumping and the continuous walls of the channel were fabricated \modf{in-situ} simultaneously. One could, however, incorporate an active pump to feed a preexisting channel. In this system, we observe that the rectified flow speed decays downstream inside the channel, but it maintains the directionality, which is imposed by the active pumps (Fig. \ref{Fig:channel}b). In this example, the length of the channel is similar to that of the active pump system, and is limited by the field of view in our current experimental setup. We have confirmed the experimental results using numerical simulations with a channel much longer than the size of the AN pumping unit (Fig. \ref{Fig:channel}c). We find that the average velocity in the center of the channel stabilizes to the value at the output from the pump (Fig. \ref{Fig:channel}c). In these simulations, the channel is just slightly wider than the flow lane of the injecting pump system. For wider channels, we observe a decay of the speed, similar to the one observed in experiments (Fig. \ref{Fig:channel}d). \modf{The effect of channel width on the sustained horizontal velocity after injection is further explored using simulations in Fig. \ref{SFig:injection_sim}. For the same channels, and in the absence of active injection, the average horizontal velocity fluctuates around zero.} 

The pumped flows can be sustained within long channels, as demonstrated with larger scale simulations in Fig. \ref{SFig:channel}. Results are different if the output of the active pump system is directed towards an arrangement of non-pumping obstacles, such as an array of circular columns (Fig. \ref{SFig:circles}), where the permeable channel walls lead to a quick randomization of the velocity.

%Finally, we designed a system that sustained a steady active flow circulation along channels of arbitrary shape. 
Our strategy allows to impose a steady active flow circulation along channels of arbitrary shape. As an example, in Fig. \ref{Fig:channel}e for experiments, and in Fig. \ref{Fig:channel}f for the corresponding simulations, we show active flows generated by a large square loop of active pump elements.  Interestingly, the active flows in the region surrounded by the loop of active pumps are no longer isotropic, which provides us with a tool to generate alternative flow patterns normally not found in active nematics. In this example, the average counterclockwise flow field created by the ring of active pumps develops a steady counterclockwise flow in the central region surrounded by triangular obstacles. We have monitored the evolution of the flow circulation along the center of the square channel (Fig. \ref{SFig:circulation_sim}), and have observed that a steady state is established after a brief transient. \modf{Besides the versatility demonstrated by these examples, our approach offers real time reconfigurability of the fluidic devices. In Movie S7 (also simulations in Movie S8) we present experiments where a simple array of active pumps is imprinted, organizing a rightwards average flow inside the virtual channels. Then, we imprint larger triangular obstacles on top of the original ones, this time rotated 180 degrees. After a short transient, the flow is reversed by the new array of pumps. This can be observed in Fig. \ref{Fig:channel}g and \ref{Fig:channel}h, where the evolution of the average horizontal velocity is represented for experiments and simulations, respectively.} 

\section{Discussion}

Achieving flow control is crucial in lab-on-a-chip devices. Fluids are driven by pressure gradients that propagate through miniaturized channels and valves with flows that obey the fundamental laws of hydraulics, namely, mass conservation and fluxes that are proportional to local pressure gradients. The strategies developed for these microfluidic devices cannot be directly adapted to harness the power of active fluids, where flows are self-sustained, often inherently disordered, and evolving following modified laws of hydraulics, as it has been recently demonstrated \cite{jorge2024,keogh2024}.

Amidst this complexity, we have demonstrated here a versatile protocol to locally control the directionality of the flows of a microtubule/kinesin AN. By incorporating photopolymerizable reactants in the aqueous medium, we have polymerized lattices of columns with triangular cross section, which force an adaptation of the surrounding active flows, breaking the fore-aft symmetry and transforming the usual isotropic active turbulence regime into ordered arrays of counter-rotating vortices that organize the AN flows along virtual channels, as material flux is allowed across them. Combining experiments and numerical simulations, we have analyzed both the orientational field, the flow velocity, and hydrostatic pressure profiles. We have identified a pattern in the time-averaged defect arrangement that demonstrates that the physical mechanism at the origin of active pumping is the modification of the defect dynamics due to the interaction of the active nematic with the inclusions. Our study has also demonstrated that the new structures behave as classical hydraulic pumps when a small number of elements are created sequentially, with a velocity and hydrostatic pressure that are proportional to the number of pumping elements. 

The demonstrated active pump systems can be implemented to tailor complex active flow patterns, allowing to select both the direction and magnitude of the self-sustained flows, and can also be used to design arbitrary mixing patterns of advected particles. Overall, the proposed strategy harnesses the self-sustained flows generated by active nematics into active fluid pumping, possessing the necessary qualities for autonomous transport and mixing in future bioenabled microfluidic devices.

We have also demonstrated that a suitable design strategy will allow to force active fluids into developing flow patterns that would not normally occur as the result of their self-organization, such as large scale steady flow circulation. In summary, active nematic pumps can bring us closer to effectively harnessing self-sustained active fluid layers for the development of bio-inspired micromachines. 

Finally, it is also worth stressing that the use of biocompatible fabrication protocols, as proposed here, ensures the general applicability of these results to different active fluids, both synthetic and of biological origin.

\section*{Acknowledgements}
We thank M. Pons, A. LeRoux, and G. Iruela for their assistance in the expression of motor proteins.  I.V.-C. acknowledges funding from Generalitat de Catalunya though a FI-2020 PhD. Fellowship. P.G. acknowledges support from the European Union’s Horizon Europe Research and Innovation Programme through the Marie Sklodowska-Curie Actions (grant agreement 101065794). I.V.-C., J.I.-M., and F.S. acknowledge funding from MICIU/AEI/10.13039/501100011033 (Grant No. PID2022-137713NB-C21). I.V.-C. and J.I.-M. acknowledge funding from MICIU/AEI/10.13039/501100011033 (Grant No. PDC2022-133625-I00).
R. C. and M.T.G. acknowledge financial support from the Portuguese Foundation for Science and Technology (FCT) under the contracts: PTDC/FISMAC/5689/2020 (DOI 10.54499/PTDC/FIS-MAC/5689/2020), EXPL/FIS-MAC/0406/2021 (DOI 10.54499/EXPL/FIS-MAC/0406/2021), UIDB/00618/2020 (DOI 10.54499/UIDB/00618/2020), UIDP/00618/2020 (DOI 10.54499/UIDP/00618/2020), DL 57/2016/CP1479/CT0057 (DOI 10.54499/DL57/2016/CP1479/CT0057) and 2023.10412.CPCA.A2. The authors are indebted to the Brandeis University MRSEC Biosynthesis facility, supported by NSF MRSEC 2011846, for providing the tubulin.\\

\textbf{Author contribution}. I.V.-C. performed most experiments and data analysis, with collaboration from P. G.; M. V. performed and analyzed some of the experiments; R.C.V.C. performed the numerical simulations, with collaboration from M.T.G.; J. I.-M. and F. S. supervised the research; J. I.-M. wrote the manuscript with collaboration from all authors.\\

\textbf{Data availability}. All study data are included in the article and/or supporting information. The data can also be accessed at: doi: 10.17605/OSF.IO/N8G4C. \cite{OSF}\\

\textbf{Competing interests}. The authors declare no competing interests.

\section*{Methods}

\subsection{Experiments}

\subsubsection{Active gel preparation}
Microtubules (MTs) were polymerized from heterodimeric ($\alpha$,$\beta$)-tubulin from bovine brain (Brandeis University Biological Materials Facility). Tubulin (8 mg mL\textsuperscript{-1}) was incubated at 37 °C for 30 min in aqueous M2B buffer (80 mM Pipes (piperazine-N,N'-bis(2-ethanesulfonic acid)), 1 mM EGTA (ethylene glycol-bis($\beta$-aminoethyl ether)-N,N,N',N'-tetraacetic acid), 2 mM MgCl\textsubscript{2}) (Sigma; P1851, E3889 and M4880, respectively) prepared with Milli-Q water and supplemented with the antioxidant agent dithiothrethiol (DTT) (Sigma 43815) and GMPCPP (guanosine-5'-{[}($\alpha$,$\beta$)-methyleno{]}triphosphate) (Jena Biosciences, NU-405) up to a concentration of 1 mM and 0.6 mM, respectively. GMPCPP is a non-hydrolysable guanosine triphosphate (GTP) analogue that favours the formation of stable MTs, obtaining high-density suspensions of short MTs (1-2 $\mu$m). 3\% of the total tubulin concentration was labelled with Alexa 647 to allow their observation by means of fluorescence microscopy. Afterward, the mixture was annealed at room temperature for 4 h and kept at -80 °C until use.

\emph{Drosophila melanogaster} heavy-chain kinesin-1 truncated at residue 401 fused to biotin carboxyl carrier protein (BCCP), and labeled with six histidine tags (K401-BCCP-6His) was expressed in BL21(DE3)pLysS competent cells using the plasmid WC2 from the Gelles Laboratory (Brandeis University) and purified with a nickel column.After dialysis against 500 mM imidazole aqueous buffer, kinesin concentration was estimated by means of absorption spectroscopy.  Finally, the kinesin was stored in a 40\% (wt/vol) aqueous sucrose solution at -80 °C until use.

Kinesin motor clusters were prepared by incubating on ice for 30 min the motor protins with streptavidin (Invitrogen; 43-4301) at a specific ratio 2:1. MTs were mixed with the motor clusters, ATP (adenosine 5'-triphosphate) (Sigma; A2383) to fuel the motors, MTs, and the depleting agent PEG (poly(ethylene glycol)) (20 kDa, Sigma; 95172). An enzymatic ATP regenerator system was added to ensure the duration of activity for several hours, it consisted on phosphoenolpyruvate (PEP) (Sigma; P7127) that fuelled the enzyme pyruvate kinase/lactate dehydrogenase (PKLDH) (Sigma; P0294) to convert ADP (adenosine 5'-diphosphate) back to ATP. The mixture also contained several antioxidants to avoid photobleaching during the fluorescence imaging: DTT, trolox (6-hydroxy-2,5,7,8-tetramethylchroman-2-carboxylic acid) (Sigma; 238813), glucose oxidase (Sigma; GT141), catalase (Sigma; C40), and glucose (Sigma; G7021). Furthermore, the constituents of the photopolymerizing hydrogel were added to the active gel. This photopolymerizing hydrogel was composed of a photoinitiator, LAP (lithium phenyl-2,4,6-trimethylbenzoylphosphinate) (TCI; L0290), and a monomer, a multiarmed PEG derivative with acrylate groups at each terminal (4-ArmPEG-Acrylate 5 kDa) (Biochempeg; A44009-5k). Additionally, Dragon Green (DG) fluorescent polystyrene particles of diameter 1 $\mu$m (Bang Laboratories; FSDG004) were added to extract the flow field of the active material. Table \ref{table:Conc} shows the final concentrations of each reagent used in the preparation of the active material.

\begin{table}[h]
    \begin{tabular}{|l|l|c|l|}
    \hline
        \textbf{Compound} & \textbf{Buffer} & \textbf{Final Conc.} & \textbf{Units}  \\ \hline
        Microtubules & Original & 1.33 & mg/mL  \\ \hline
        Kinesin & Original & 0.32 & $\mu$M  \\ \hline
        Streptavidin & M2B & 0.01 & mg/mL  \\ \hline
        ATP & M2B & 300 & $\mu$M  \\ \hline
        PEG (20 kDa) & M2B & 1.6 & \% w/v  \\ \hline
        PEP & M2B & 27 & mM  \\ \hline
        MgCl$_2$ & M2B & 3.3 & mM  \\ \hline
        DTT & M2B & 5.8 & mM  \\ \hline
        Trolox & Phosphate & 2.1 & mM  \\ \hline
        Catalase & Phosphate & 0.04 & mg/mL  \\ \hline
        Glucose & Phosphate & 3.20 & mg/mL  \\ \hline
        Glucose Oxidase & Phosphate & 0.21 & mg/mL  \\ \hline
        PK & Original & 27 & u/mL  \\ \hline
        LAP & Mixture & 0.226 & \% w/v  \\ \hline
        4PEG5k & Mixture & 4.52 & \% w/v  \\ \hline
        DG Particles & Original & 1 & \% v/v  \\  \hline
    \end{tabular}
    \caption{Composition of all stock solutions (including buffer used for their preparation), and concentration of the different species in the final mixture. Acronyms used in this table are: PEG (Poly-ethylene glycol); PEP (Phosphoenol pyruvate); ATP (Adenosin triphosphate); PK (Pyruvate Kinase); DTT (1,4-dithiothreitol); 4PEG5k (4-ArmPEG-Acrylate 5 kDa); LAP (lithium phenyl-2,4,6-trimethylbenzoylphosphinate). The two latter compounds where directly incorporated to the active mixture. M2B buffer: 80 mM PIPES (piperazine-N,N'-bis(2-ethanesulfonic acid)) pH 6.8, 2 mM MgCl$_2$ 1 mM EGTA (egtazic acid). Phosphate buffer: 20 mM Phosphate buffer (6.68 mM KH$_2$PO$_4$, 12.32 mM K$_2$HPO$_4$) pH 7.2;  Original: species is obtained already dissolved in its custom buffer. }
    \label{table:Conc}
\end{table}

\subsubsection{Active nematic cell}
Flow cells were prepared with a bioinert and superhydrophilic polyacrylamide-coated glass and a hydrophobic aquapel-coated glass. A 50 µm height double-sided tape was used as a spacer, leaving a channel of $\sim$4 mm width. The cell was filled first with fluorinated oil (HFE7500; Fluorochem; 051243) which contains a 2\% of a fluorosurfactant (008 Fluorosurfactant; RanBiotechnologies) to get a surfactant-decorated interface. Afterwards, the active material was introduced into the cell by capillarity and the cell was sealed using petroleum jelly to avoid evaporation.

\subsubsection{Imaging and photopatterning}
The observation of the active material was done by means of fluorescence microscopy. We used a Nikon Eclipse Ti2-e equipped with a white LED source (Thorlabs MWWHLP1), a Cy5 filter cube (Edmund Optics) and a FITR filter cube (Thorlabs, TLV-TE2000-FITC). Images were captured using an Andor Zyla 4.2 Plus camera or a QImaging ExiBlue cooled CCD camera operated with ImageJ $\mu$-Manager open-source software.

The fluorescence microscope was modified in order to incorporate a Ti Light Crafter 4500 DLP development modules (EKB Technologies, Ltd.) equipped with a high 2 W 385 nm LED source in one case. Projected patterns are incorporated into the light path of the inverted microscope by means of a collimating lens (f = +150 mm) and a 505 nm dichroic mirror (Thorlabs DMLP505R), and are focused on the sample by means of the microscope objectives, reaching a lateral resolution up to a few microns. The DMD projector is connected as an external monitor to a computer, thus enabling real time control of the projected patterns using MS-PowerPoint slides. Hydrogel objects were polymerized using the Nikon CFI Plan Fluor DLL x10 objective (N.A. = 0.3), resulting in a light power density of 1.6 W cm\textsuperscript{-2}), and a irradiation time of 2 seconds. The imprinted patterns are currently confined within the field of view, which is around $850\,\mu$m$\times650\,\mu$m in the used optical setup. The studied active pump systems are in all cases surrounded by extended regions, at least 1 mm in all directions, where the unconstrained active nematic is in the active turbulent regime.

\subsubsection{Flow field analysis and particle tracking}
The flow field of the active material was extracted by performing particle image velocimetry directly on the fluorescence images using the public domain MatLab App PIVLab. To study the transport of advected colloidal particles, we have included a small amount of $1\,\mu$m hydrophilic fluorescent polystyrene microparticles (Dragon Green, FCDG006, Bangs Labs), and tracked their motion from fluorescence images using the plugin Particle Tracker 2D/3D in FIJI.

\subsection{Orientational field analysis and defect detection}
The local active nematic tensor field was obtained using a coherence-enhanced diffusion filtering (CEDF) MatLab code \cite{Ellis2020} applied on fluorescence images of the AN layer. This analysis yields the local scalar order parameter and director field, and allows to detect the topological defects. The local topological charge density, $\sigma$, is obtained from the director field, $n_i$ by defining the tensor $q_{ij}=n_in_j-1/2\delta_{ij}$ and $\sigma=-\frac{1}{\pi} \left( \frac{\partial q_{11}}{\partial x} \frac{\partial q_{12}}{\partial y} + \frac{\partial q_{12}}{\partial x} \frac{\partial q_{22}}{\partial y} - \frac{\partial q_{12}}{\partial x} \frac{\partial q_{11}}{\partial y} - \frac{\partial q_{22}}{\partial x} \frac{\partial q_{12}}{\partial y} \right)$.

\subsection{Simulations}

We model the active nematics using a 2D continuum hydrodynamic theory which is summarized in this section. The director field  $\mathbf{n}(\mathbf{x})$ gives the average orientation of the filaments in the AN at the position $\mathbf{x}$ while the degree of order is given by the scalar order parameter $S$, which is $S=0$ in the isotropic phase and has a finite value $S_N$ in the nematic phase. Because the filaments have a head-tail symmetry, a tensor order parameter is used: $Q_{\alpha \beta} = 2 S \left ( n_{\alpha} n_{\beta} - \delta_{\alpha \beta} /2\right )$. \textcolor{blue}{In the free energy density, the bulk ordering terms are not considered. This corresponds to placing the passive nematic at the instability of the isotropic phase, the isotropic spinodal. At this point the passive system is unstable to nematic fluctuations. Our motivation for this choice of parameters arises from experimental observations in microtubule-kinesin mixtures, where the nematic order emerges driven by activity alone. While including bulk terms modifies the stability of the passive system, these terms are not expected to affect the dynamical steady state qualitatively, as activity drives both the passive ordered state and the disordered state considered here linearly unstable with a zero threshold for large systems. A similar approach has been considered in related studies~\cite{Thampi_2015, Hardouin2019, COELHO2024100309}}. The free energy density includes an elastic term responsible for penalizing distortions in the order parameter, $f_{E}= K\left(\partial_{\gamma} Q_{\alpha \beta}\right)^{2}/2 $, where $K$ is the elastic constant, and an anchoring term applied only at the solid surfaces, $f_{W}= W(Q_{\alpha \beta}-Q_{\alpha \beta}^{0})^{2}/2$, which penalises deviations from the surface-preferred order parameter $Q_{\alpha \beta}^{0}$ with strength $W$. The total free energy is the sum of these two terms integrated in the entire area: $\mathcal{F} = \int f d^2 x $, where $f=f_E+f_W$ at the solid surface and $f=f_E$ elsewhere. The dynamics is governed by the Beris-Edwards (for the tensor order parameter $Q_{\alpha\beta}$), continuity and Navier-Stokes equations (for the velocity field $\mathbf{u}$), respectively~\cite{beris1994thermodynamics, PhysRevLett.89.058101,doostmohammadi2018active}:
\begin{align}
  &\partial _t Q_{\alpha \beta} + u _\gamma \partial _\gamma Q_{\alpha \beta} - S_{\alpha \beta} = \Gamma H_{\alpha\beta} ,   \label{beris-edwards-eq}\\
  &\partial_\alpha  u_\alpha =0.   \label{contintuity-eq}\\
  &\rho \partial _t u_\alpha + \rho u_\beta \partial_\beta u_{\alpha} = -\partial_\alpha p  + 2 \eta \partial_\beta D_{\alpha\beta}  -\zeta \partial_ \beta Q_{\alpha\beta}.\label{navier-stokes-eq}
\end{align}
In Eq.~\eqref{beris-edwards-eq}, $\Gamma$ stands for the rotational diffusivity, $H_{\alpha\beta}$ for the molecular field: $ H_{\alpha\beta} = -\frac{\delta \mathcal{F}}{\delta Q_{\alpha\beta}} + \frac{\delta_{\alpha\beta}}{2} \Tr \left( \frac{\delta \mathcal{F}}{\delta Q_{\gamma \epsilon}} \right)$
and $S_{\alpha \beta}$ for the co-rotational term: $ S_{\alpha \beta} = ( \xi D_{\alpha \gamma} + W_{\alpha \gamma})\left(Q_{\beta\gamma} + \delta_{\beta\gamma}/2 \right)
+\left( Q_{\alpha\gamma}+\delta_{\alpha\gamma}/2 \right)(\xi D_{\gamma\beta}-W_{\gamma\beta})
-2\xi\left( Q_{\alpha\beta}+\delta_{\alpha\beta}/2  \right)(Q_{\gamma\epsilon} \partial _\gamma u_\epsilon)$, where the vorticity and shear rate are $W_{\alpha\beta} = (\partial _\beta u_\alpha - \partial _\alpha u_\beta )/2$ and $D_{\alpha\beta} = (\partial _\beta u_\alpha + \partial _\alpha u_\beta )/2$. The aligning parameter $\xi$ controls the relative importance of the vorticity and shear rate and it is related to the shape of the particles that compose the nematics: rod-like particles have $\xi>0$, while disk-like particles have $\xi<0$. We consider rod-like particles (filaments) in the flow-aligning regime. In Eq.~\eqref{navier-stokes-eq}, $\rho$ is the fluid density, $p$ is the fluid pressure and $\eta$ is the absolute viscosity. The activity level is controlled by the parameter $\zeta$, which is positive for extensile systems as those in this work. The equations described above are a simplified version of a standard 2D model for active nematics~\cite{Thampi_2015, C9SM00859D, doostmohammadi2018active}. Because the active stresses are typically much larger than the passive ones~\cite{velez2024}, we neglected the passive stress in Eq.~\eqref{navier-stokes-eq} as in previous works~\cite{Hardouin2019, PhysRevE.106.014705}. While the director field can point out of the plane in 3D models under certain conditions~\cite{Nejad2022}, this behaviour is not observed in the present experiments. Therefore, the 2D model appropriately describes the reported phenomena.

A hybrid numerical method is used in the simulations. Equation~\eqref{beris-edwards-eq} is solved using a predictor-corrector finite-differences method while Eqs.~\eqref{contintuity-eq} and~\eqref{navier-stokes-eq} are recovered in the macroscopic limit using the lattice Boltzmann method~\cite{marenduzzo2007steady, C9SM00859D}. At the solid obstacle surfaces, we use planar weak anchoring and no-slip boundary conditions in line with known experimental conditions \cite{hardouin2020}, by means of half-way bounce-back boundary conditions. We have verified that, even at zero anchoring strength, the directors align mostly parallel to the obstacle walls. Incrementing the anchoring strength only decreases the average speed. Similarly, we have tested numerical simulations with free-slip boundary conditions of the velocity field on the obstacle walls, finding the same qualitative results. The reported pumping phenomenon, therefore, is robust with respect to the interactions between the active nematic material and the obstacle walls.
Simulations are either performed within a closed box, or with periodic boundary conditions in all directions. For closed boxes, the size of the simulation box is adapted to the obstacle system under study, allowing a sufficiently wide surrounding layer of turbulent active nematic. We have verified that this choice of boundary conditions has no relevant impact on the qualitative performance or in the velocity or pressure profiles within the studied pumping systems. 

The initial conditions are the fluid at rest and directors mostly aligned with the horizontal direction with random fluctuations of $\pm 2 ^\circ$. Different seeds are used to generate random initial conditions. The parameters used in the simulations are as follows (except where otherwise stated) in lattice units, in which the reference density, lattice spacing and time step are equal to one: absolute viscosity $\eta=6.67$, elastic constant $K=0.015$, aligning parameter $\xi=0.9$, activity $\zeta=0.01$, fluid density $\rho=40$, anchoring strength $W=0.002$, rotational diffusivity $\Gamma=0.4$, obstacle size $H=70$ and size of the square unit cell $L=128$.

The results are shown in reduced units, which are calculated considering three quantities in lattice units: the average density $\rho_0=40$, the average velocity of the free active nematics (without obstacles) $v_r\approx0.0105$ and the average distance between defects for the free active nematics $D_{def}\approx56$. All other reference quantities are calculated using these three.

The tracers are simulated as massless particles in the overdamped regime that follow the fluid velocity and do not collide among themselves. Thus, the tracer's position is calculated as $\mathbf{x}_i(t+\Delta t) = \mathbf{x}_i(t) + \mathbf{u}(\mathbf{x}) \Delta t $, where $i$ is the index that identifies each tracer and $\Delta t$ is the times step from the lattice Boltzmann simulation.

\bibliography{biblio}% Produces the bibliography via BibTeX.

\end{document}